\documentclass[aps,twocolumn,showpacs,superscriptaddress,showkeys,prl,10pt,groupedaddress]{revtex4-1}
\usepackage{dcolumn}   
\usepackage{bm}      
\usepackage{xcolor}
\usepackage[utf8]{inputenc}
\usepackage{amsmath, amssymb,slashed}
\usepackage{float}
\usepackage{graphicx}
\usepackage[english]{babel}
\usepackage{hyperref}
\usepackage{tikz}

\usepackage{comment}
\usepackage{silence}

\newcommand{\beq}{\begin{equation}}
\newcommand{\eeq}{\end{equation}}                     
\newcommand{\beqa}{\begin{eqnarray}}
\newcommand{\eeqa}{\end{eqnarray}}

\usetikzlibrary{decorations.pathmorphing}
\usetikzlibrary{decorations.markings}
\usetikzlibrary{mindmap}
\usepgflibrary[shapes.arrows]
 \definecolor{darkgreen}{RGB}{34,139,34}

\begin{document}

\title{Reactor and Atmospheric Neutrino Mixing Angles' Correlation as a Probe for New Physics}

\author{Pedro Pasquini}
\address{Instituto de F\'isica Gleb Wataghin - Universidade Estadual de Campinas -UNICAMP, {13083-859}, Campinas SP, Brazil}

\begin{abstract}
 We performed a simulation on the DUNE experiment to probe the capability of future neutrino long-baseline experiments' ability to constrain the parameter space of high-energy models by using the correlation between the atmospheric and reactor mixing angles. As an example, we took the Tetrahedral Flavour Symmetry model, which predicts a strong relation between the non-zero value of $\theta_{13}$ and deviation of $\theta_{23}$ from the maximality. We show that in this case, the model can realistically be excluded in more than $3\sigma$ for most of the parameter space. We also study the octant degeneracy at DUNE and this impact on the sensitivity of such models.
\end{abstract}

\maketitle
\section{\label{sec:intro} Introduction}
The future scenario of experimental neutrino physics predicts an unprecedented precision in the measurement of all the neutrino mixing parameters. In special, the atmospheric angle and the CP phase are the two most unknown and are expected to be well measured at future long-baseline experiments.

It was shown in~\cite{Pasquini:2016kwk} that one can use such precision in order to tightly constrain the allowed parameter space of high-energy models, as long as they provide robust relations between the neutrino mixing parameters. In special, the correlation of the $\theta_{23}$ angle and the $\delta_{\rm CP}$ can be used in oscillation experiments to distinguish it from the usual 3-neutrino standard paradigm up to 5$\sigma$ C. L.~\cite{Chatterjee:2017xkb,Chatterjee:2017ilf}. Also, the phenomenology of the reactor angle in {\it Residual $ Z_2$ Symmetry models} was studied in~\cite{Ge:2011qn,Hanlon:2013ska} in the context of the prediction of $\delta_{\rm CP}$.
In this letter, we show that we can go even further and use the correlation between the atmospheric and the reactor angles as a probe for high-energy physics. 

As an example, we took the Tetrahedral Flavour Symmetry model (TFSM) presented in~\cite{Eby:2008uc} in order to obtain the regions of the parameter space that can be excluded by the DUNE experiment~\cite{Acciarri:2015uup}.

We also found that DUNE experiment has a degeneracy for lower values of the $\sin^2\theta_{13}$ which does not present any consequences for standard 3-Neutrino oscillations, but partly reduces the test capability of such correlations.
\section{\label{sec:model} The Tetrahedral Symmetry Model}
The Tetrahedral Symmetry model~\cite{Frampton:2008bz} consists on the addition to the standard model's gauge group $G$, two discrete flavor symmetries described by the Binary Tetrahedral group $T_4$ and a $Z_2$ symmetry, that is: $G_{\rm tot}=G\times T_4\times Z_2$. 

The particle content of the model consists of the usual standard model particles, plus two singlet scalars to accommodate the quark masses, two triplets scalars to accommodate both charged and neutral fermion masses and three singlets neutral fermions.

The neutrinos acquire masses by a see-saw mechanism through the 3 extra singlets neutral fermions, $N^{i}_R$, $i=1,2,3$ and one of the scalar triplet, $H_3=(H_3^1,H_3^2,H_3^3)$. The masses arise from a spontaneous symmetry break of the scalar potential when one makes the change $H_3\rightarrow h_3+\vec{v}$ where $h_3$ is the triplet after symmetry break and $\vec{v}=v(1,-2,1)$ the vacuum expectation value. In this configuration, the model predicts a Tri-Bi-maximal (TBM) mixing matrix,

The TBM matrix is now excluded as it requires a zero reactor mixing angle. Ne\-ver\-the\-less, it was shown in~\cite{Eby:2008uc}, that by a small shift the vacuum alignment of $H_3$, non-zero value of the reactor angle can be obtained. Such values are directly related to the atmospheric mixing by inducing a connection between the non-zero value of $\theta_{13}$ and $\theta_{23}$,
\begin{equation}\label{eq:contrain}
 \theta_{13}=\sqrt{2}\left|\frac{\pi}{4}-\theta_{23}\right|.
\end{equation}
This is a very powerful prediction and relates the non-zero $\theta_{13}$ to deviations of $\theta_{23}$ from the maximality. We will show that the constraint obtained from Eq.~\ref{eq:contrain} can be used to probe the model in long-baseline experiments. A discussion and extension of $A_4$ based models and residual $Z_2$ due to spontaneous breaking can be found in~\cite{King:2014nza,Dicus:2010yu}.

\section{\label{sec:res} Probing the Model}
In order to show the capability of future neutrino experiments to distinguish the standard neutrino paradigm from constrained models, we simulated the DUNE experiment with the configuration described in~\cite{Acciarri:2015uup} using the Globes software~\cite{Huber:2004ka,Huber:2007ji}. 

The DUNE experiment consists of a muon/electron (anti-)neutrino beam from the Fermilab facility that travels through the earth mantle for 1300 km reaching a 40 kt liquid argon detector at Ash River, Minnesota. The beam's energy peaks around 2 GeV and the detector is optimized to measure the neutrino disappearance rate for $\mu$-(anti)neutrino and the e-(anti)neutrino appearance. The precision achieved on $\theta_{23}$ is related to the precision of the $\theta_{13}$ angle~\cite{Chatterjee:2017irl} and could potentially reach $6\%$. 

\begin{figure}[!h]
\centering
\includegraphics[scale=.34]{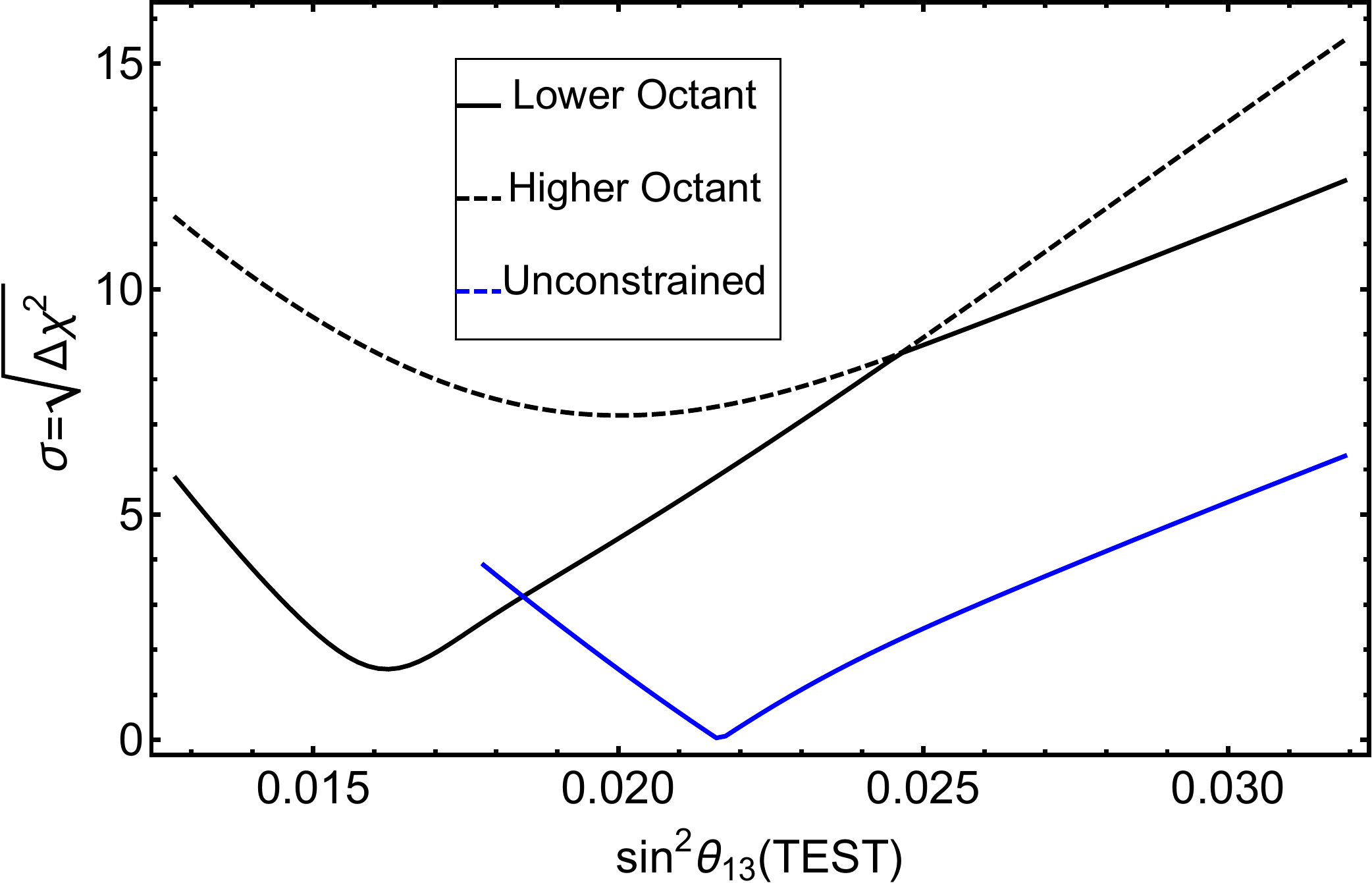}
\caption{\label{fig:plot1} $\chi^2$ as a function of the test value of $\theta_{13}$. The blue curve represents the standard paradigm while the Black curves is the prediction by the tetahedral model, the continuos line corresponds to the lower octant while the dashed line to the higher octant. In this analysis the true value considered for the atmospheric angle is $\sin^2\theta_{23}=0.43$. }
\end{figure}
In order to perform the simulation we assumed the usual Poissonian $\chi^2$ for each energy bin. A discussion of this method can be found in~\cite{Chatterjee:2017xkb,Huber:2004ka,Huber:2007ji}. The exclusion bands were found by minimizing the $\chi^2$ over all the free parameters. The complete $\chi^2$ is a function of two sets of parameters, the {\it true values}, which are the values choosen by nature and the {\it test values}, which are the observed by the experiment. The $\chi^2$ functions minimizes over all the free test values and is defined for each pair of true values $\sin^2\theta_{13}(TRUE)$ and $\sin^2\theta_{13}(TRUE)$ in two scenarios: (1) Unconstrained  and (2) TFSM. The free parameters in each case is described in Table~\ref{pars_tab}. The order true values are kept fixed unless stated otherwise. Finally, the $\Delta \chi^2$ is defined as,\begin{align}\label{eq:chi}
\Delta \chi^2=&\chi_2^2\left[\theta_{23}(TRUE),\theta_{13}(TRUE)\right]\\ \nonumber 
&-\chi_1^2\left[\theta_{23}(TRUE),\theta_{13}(TRUE)\right]                                                                                                                                                                                                                                                                                                                              \end{align}
where $\chi_i^2$ is the minimum $\chi^2$ for assumption $i=$(1),(2). Thus, $n\sigma$ C. L. exclusion means $\Delta \chi^2\ge n^2$. The fixed true values were taken as best fit from~\cite{deSalas:2017kay}. The  signal (background) flux uncertanty is assumed to be 5\% (10\%) and a 3.5 year neutrino/anti-neutrino mode run each.
\begin{table}[H]
\centering
 \begin{tabular}{cccc}
  Parm. & (1) Unconst. & (2) TFSM & Central Value\\ \hline \hline
  $\theta_{12}(TEST)$ &  Free & Free &$34.5^0(1)$\\
  $\theta_{13}(TEST)$ & Free &Free&$8.44^0(18)$\\
  $\theta_{23}(TEST)$ & Free & Constr. &$41^0(1.1)$ or $50.5^0(1)$ \\
  $\delta_{\rm CP}(TEST)$ & Free &Free& $1.4\pi$\\
  $\Delta m_{21}^2(TEST)$ &Free&Free& $7.56(19)\times 10^{-5}$ eV$^2$\\
  $|\Delta m_{31}^2(TEST)|$ &Free &Free&$2.55(4)\times 10^{-3}$ eV$^2$\\ \hline
 \end{tabular}
\caption{\label{pars_tab} Minimized Free or constrained parameters on the $\chi^2$ function and the central values assumed unless stated otherwise. The central values are taken from~\cite{deSalas:2017kay}.}
\end{table}

In Fig.~\ref{fig:plot1} we plotted the expected DUNE sensitivity curve as a function of $\sin^2\theta_{13}$ by taking $\sin^2\theta_{23}(TRUE)=0.43$ in three prior assmptions on the minimization procedure:
\begin{itemize}
 \item[(i)] unconstrained case (blue)
 \item[(ii)] TFSM $\theta_{23}(TEST)$ at lower octant (black) 
 \item[(iii)] TFSM $\theta_{23}(TEST)$ at higher octant (black-dashed)
\end{itemize}
In all cases the test values are minimized acording to Table~\ref{pars_tab}. The curve represents the capability of DUNE to measure the $\theta_{13}$ parameter. A $n\sigma$ Confidence Level is defined as $\chi^2(\theta)-\chi^2(\theta_{\rm min})=n^2$ for each curve. Thus, by assuming a prefered model over the other the central value is chanded considerably. In special minimum is shifted to a smaller value ($\sin^2\theta_{13}^{\rm min}=0.016$) when the TSFM is assumed as a prior. Also, the constraint forced by Eq.\ref{eq:contrain} results in an overal difference of $\chi^2_{\rm TSFM}(\theta_{\rm min})-\chi_{\rm free}(\theta_{\rm min}))=2.5$.

This happens due to the fact that a $\theta_{23}$ large deviation from the maximal is required to explain the size of the reactor angle. On the other hand, the atmospheric angle is assumed as $\sin^2\theta_{23}=0.43$, not too far away from $\pi/4$. Thus, the interplay between close to maximal $\theta_{23}$ and large $\theta_{13}$ can be used to probe the model.

\begin{figure}[!ht]
\centering
\includegraphics[scale=.34]{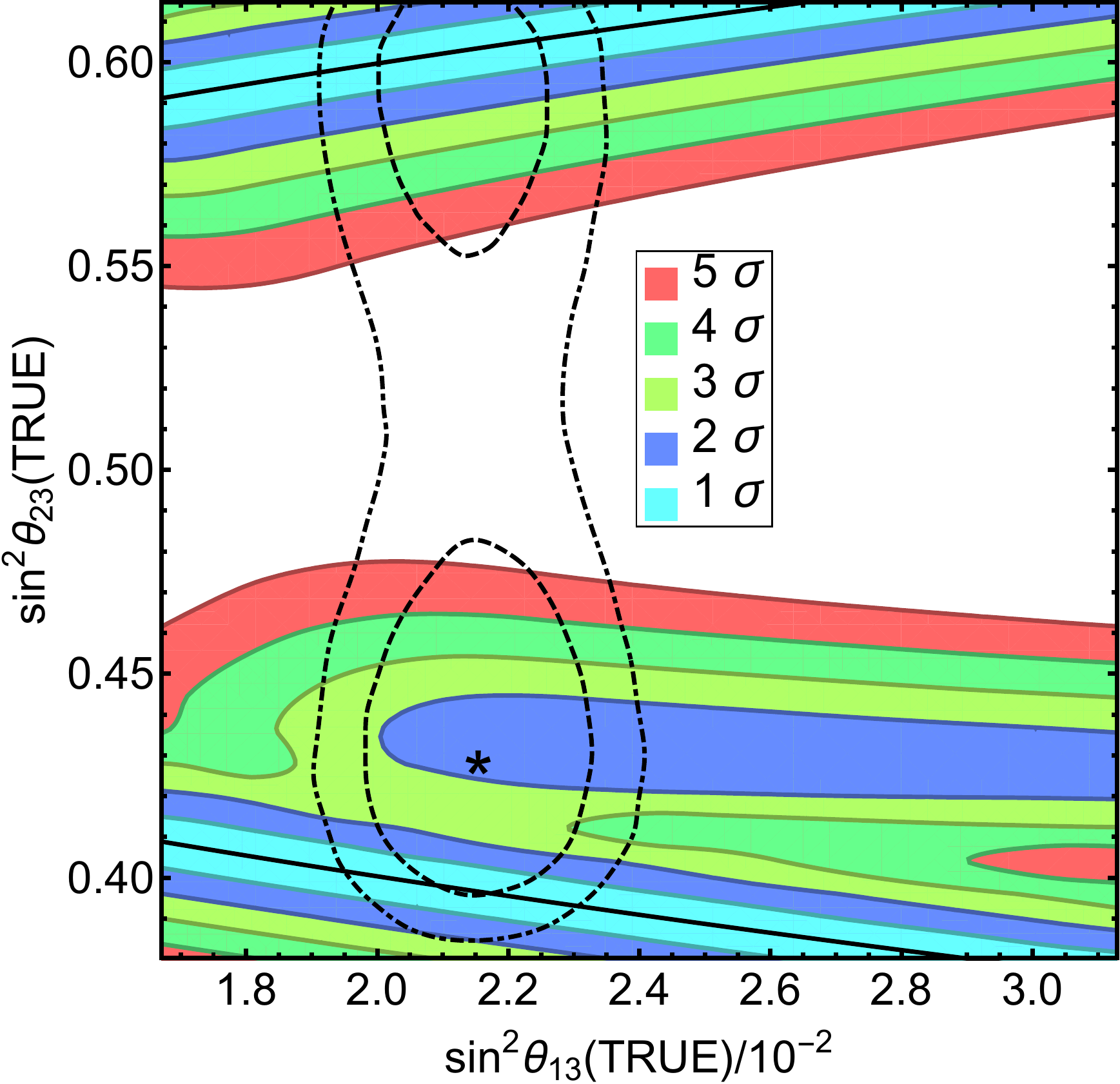}
\caption{\label{fig:plot2} Expected DUNE capability of probing the model at different confidence levels as a function of the true value of $\theta_{23}$ and $\theta_{13}$. The bands correspond to the C. L. 1 to 5 $\sigma$ while the black dashed curve represents the model prediction. The dashed (dotdashed) line corresponds to the current global fit 90\% (99\%) C. L.~\cite{deSalas:2017kay} and the star the best-fit.}
\end{figure}

In Fig.~\ref{fig:plot2} we plotted colored bands representing the 1 to 5 $\sigma$ C. L. of potential DUNE experimental exclusion. In this plot, the minimization of Eq.~\ref{eq:chi} also includes the variation of $\delta_{\rm CP}(TRUE)$ from 0 to 2$\pi$. The black line corresponds to the model prediction and the  black-dashed (black-dot-dashed) lines corresponds to the current 90\% (99\%) C. L.. Notice that the $1\sigma$ band (cyan) follows the model prediction and that the region around the maximal mixing of the atmospheric angle can be excluded with more than $5\sigma$. However, there is a $2\sigma$ degeneracy around the current best-fit point. This is due to a small $\theta_{13}$ degeneracy related to the well-known octant confusion and will be discussed in next section.

\section{\label{sec:dege}  Octant Degeneracy and Reactor Angle}
Previously it was shown that a correlation between $\theta_{13}$ and $\theta_{23}$ can be used to exclude the TFSM using the DUNE experiment alone, even though its sensibility on the reactor angle being much smaller than reactor experiments. Nevertheless, Fig.~\ref{fig:plot2} shows a clear $2\sigma$ degeneracy near the best-fit point.

\begin{figure}[!ht]
\centering
\includegraphics[scale=.215]{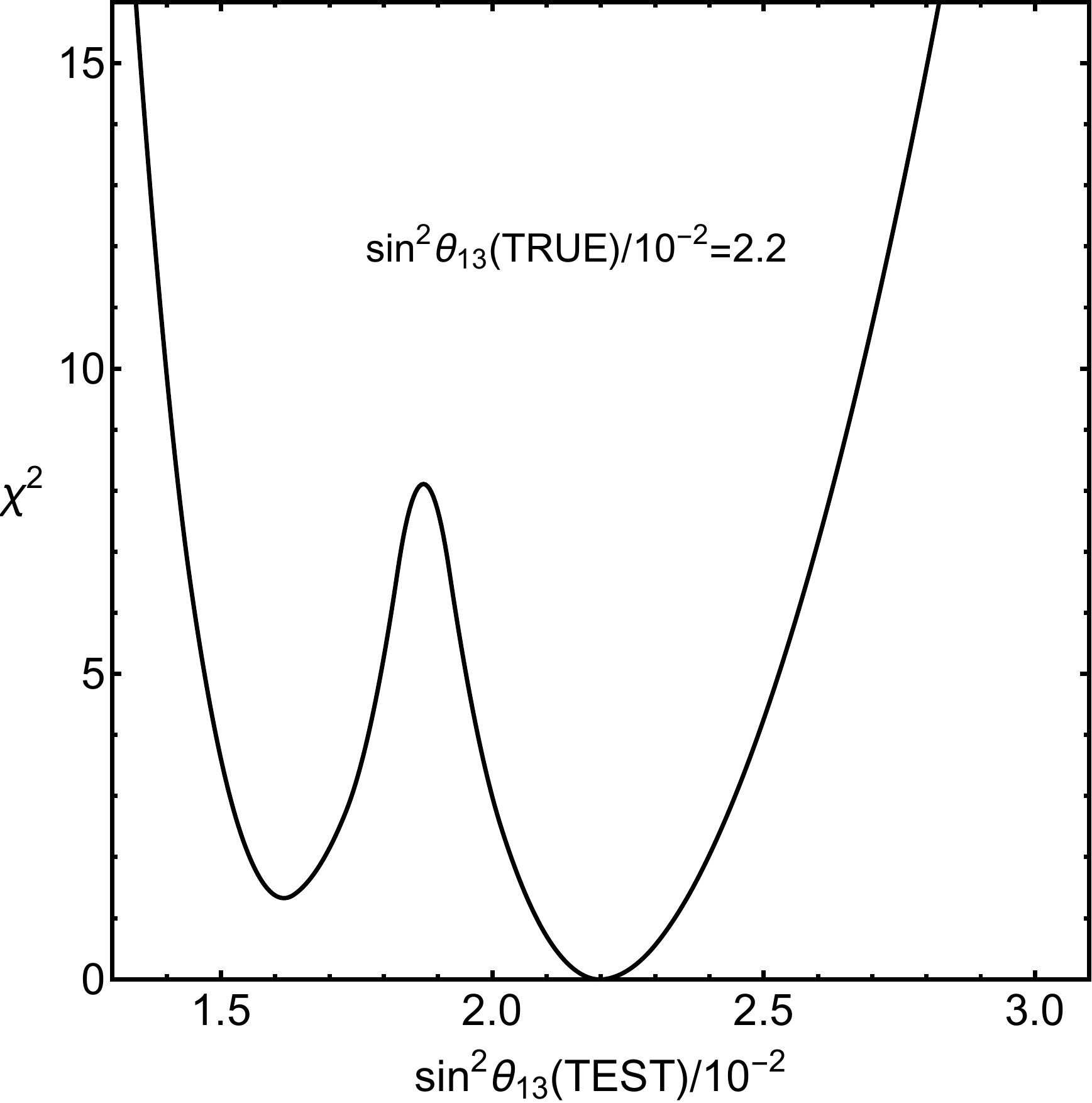}
\includegraphics[scale=.215]{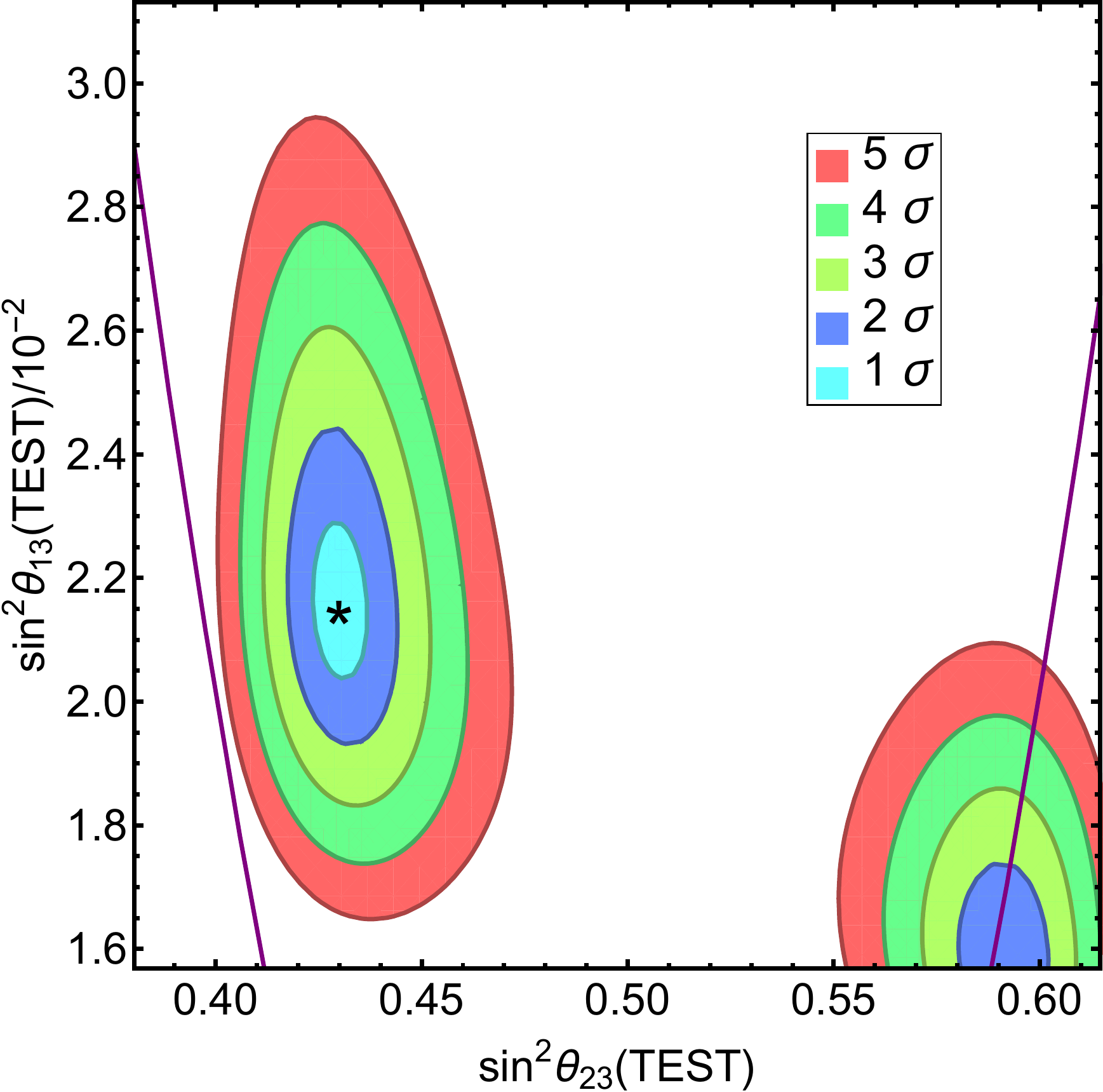}
\caption{\label{fig:plot3} Projected Precision measurement of $\sin^2\theta_{13}$ by DUNE experiment. {\bf Left:}  1D $\chi^2$ projection of $\theta_{13}$ and {\bf Right :} $\theta_{23}$ and $\theta_{13}$ correlation. The colors represents the different confidence levels and the purple line is the model prediction.}
\end{figure}

Such degeneracy is a result of the octant confusion~\cite{Minakata:2002jv}. For each pair $\left(\theta_{13}(TRUE),\theta_{23}(TRUE)\right)$ there is and eighth-fold degeneracy caused by the four {\it test} parameters $\left(\theta_{13},\theta_{23}, \delta_{\rm CP}, {\rm sing}(\Delta m_{31}^2)\right)$. Nevertheless, the correlation between $\theta_{13}$ and $\theta_{23}$ form Eq.\ref{eq:contrain} partly reduces such degeneracy as not every atmospheric angle is allowed. On the other hand, as Eq.~\ref{eq:contrain} is symmetric around $\theta_{23}=\pi/4$, one may find the second solution in the wrong octant that fits the reactor angle at $3\sigma$. This is illustrated in Fig.~\ref{fig:plot3}. The left panel presents the precision measurement of $\theta_{13}$ in the unconstrained case, assuming the true value as $\sin^2\theta_{13}/10^{-2}=2.2$. We see the usual dip due to the octant degeneracy that creates a second minimum in the $\theta_{13}(TEST)$ $\chi^2$ function. The matter is even more clear in the right panel we present the 2D confidence regions for $\theta_{13}(TEST)$ and $\theta_{23}(TEST)$. The true value is assumed at $\sin^2\theta_{23}=0.43$ and the model prediction corresponds to the purple line. Notice that even though the model completely misses the true value at the lower octant by more than $5\sigma$ it passes through a region at the higher octant at $2\sigma$. Thus, such local minimum can push down the $\chi^2$ difference between the TFSM and the usual 3-neutrino oscillation due to the wrong octant measure. This means that $\theta_{13}$ correlation cannot be used to its full potential to probe the model when considering DUNE alone.

\begin{figure}[!ht]
\centering
\includegraphics[scale=.215]{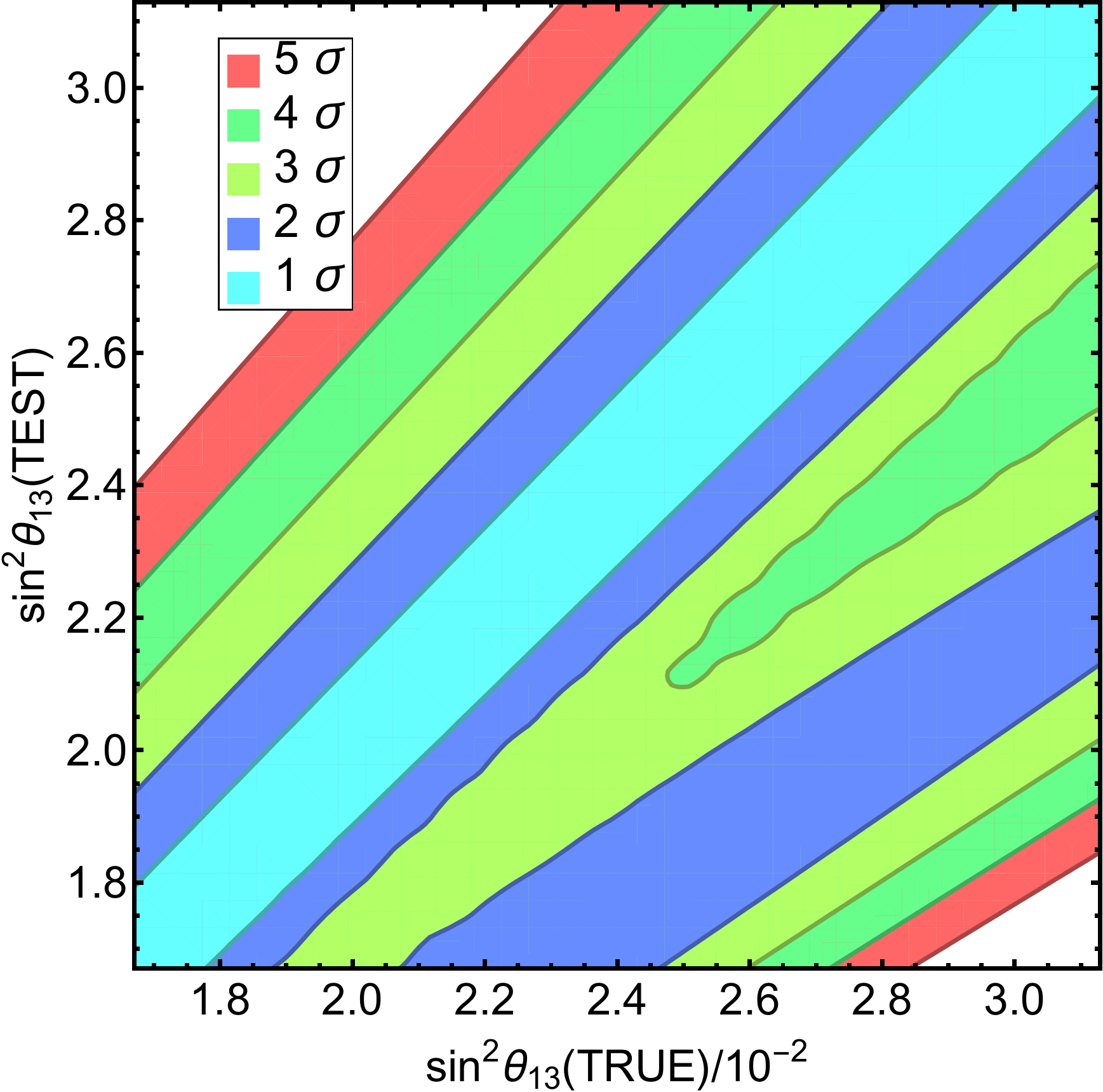}
\caption{\label{fig:plot_relation}  Projected DUNE experiment confidence Level exclusion for the $\sin^2\theta_{13}(TRUE)$-$\sin^2\theta_{13}(TEST)$ plane.}
\end{figure}

For completeness, we also present the $\theta_{13}$ precision measurement as a function of the true value of $\theta_{13}$ in the right panel of Fig.~\ref{fig:plot_relation} for different confidence levels. For each $\theta_{13}(TRUE)$ there is a smaller value of $\theta_{13}(TEST)$ that is allowed at $2\sigma$ represented by the blue band in the plot.

\section{\label{sec:gene} Generalizing Relation}
Last sections described how it is possible to use longbaseline experiments to probe the TFSM. This is possible due to the strong correlation between $\theta_{13}$ and $\theta_{23}$ in Eq.\ref{eq:contrain}. Nevertheless, we can straightfowardly generalize such relation in order to perform a model-independent analysis. If the reactor angle can be written as a function of the atmospheric angle, that is $\theta_{13}\equiv f(\theta_{23})$, one may expand it around $\theta_{23}=\pi/4$ as,

\begin{equation}
 \theta_{13}= f(\pi/4)+f'(\pi/4)\left|\frac{\pi}{4}-\theta\right|+...
\end{equation}

Thus, the TFSM is a particular case where it predicts $f(\pi/4)=\theta^0=0$ and $f'(\pi/4)=\sqrt{2}$. That means that a straightfoward generalization of Eq.~\ref{eq:contrain} is realized by substituting the proportionality factor $\sqrt{2}$ by a free factor $f=f'(\pi/4)$ as described bellow,
\begin{equation}\label{eq:contrain2}
 \theta_{13}=f\left|\frac{\pi}{4}-\theta_{23}\right|.
\end{equation}
Now, we can probe a more general, phenomenolocialy motivatede model independent relation in the DUNE experiment. Nevertheless, we showed that the octant degeneracy may interfere in the sensitivity of such relation. However, it was shown in~\cite{Minakata:2002jv,Chatterjee:2017irl} that the octant problem can be partly lifted if one knows precisely the value of $\theta_{13}$. In special, Daya-Bay may reach up to $3\%$ sensitivity in the reactor angle~\cite{Seo:2017tjb}.

\begin{figure}[!ht]
\centering
\includegraphics[scale=.4]{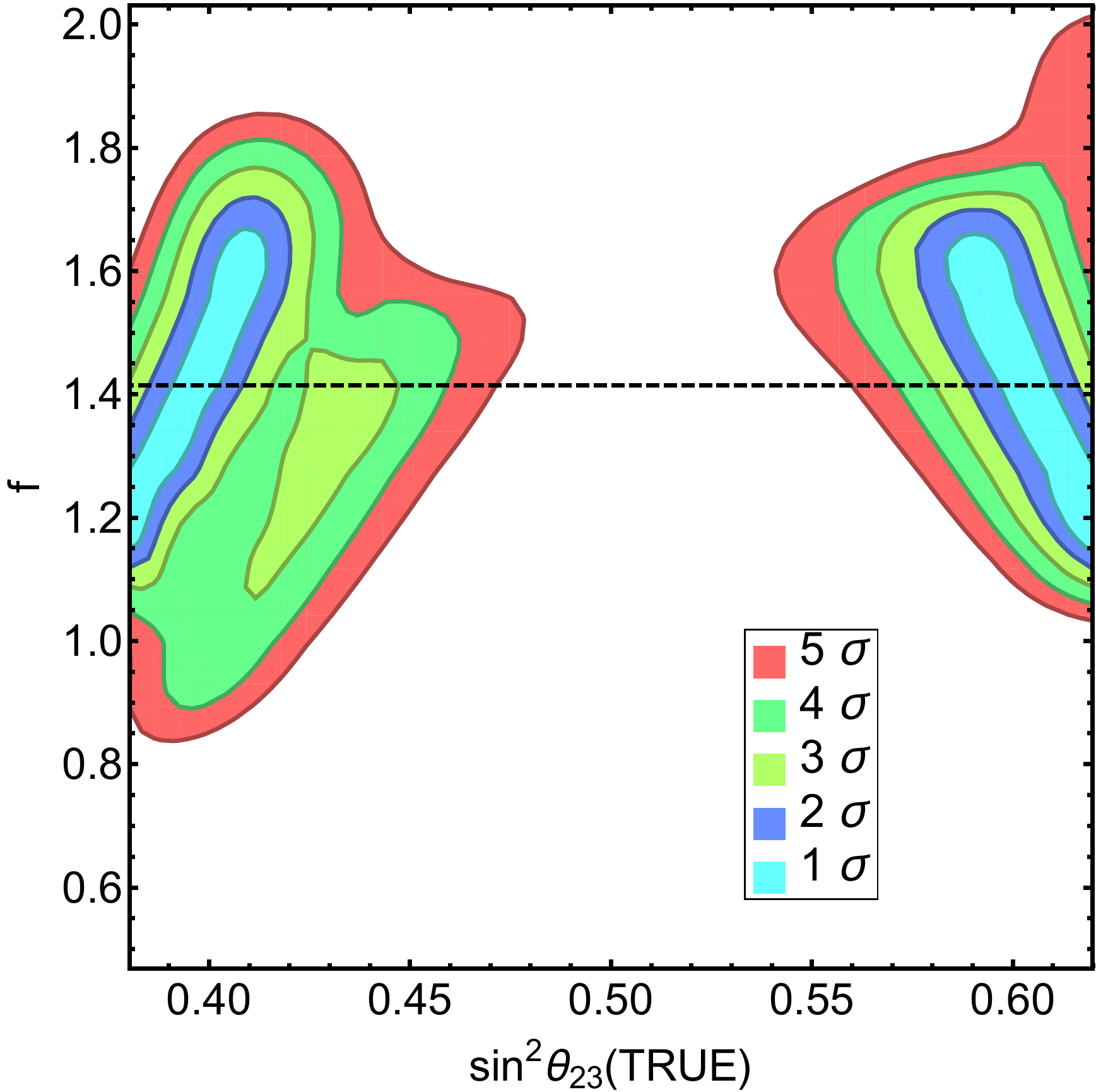}
\caption{\label{fig:plot_general} DUNE capability of excluding any relation between $\theta_{13}$ and $\theta_{23}$ that follows Eq.~\ref{eq:contrain}. The Colored curves represents 1 to 5 $\sigma$ C. L. and the dashed black curve is the Tetrahedral model.}
\end{figure}

Thus, we will combine the DUNE experiment reactors in order to evaluate the full potential of disentangling the TFSM from the standard 3$\nu$ oscillation. This is performed by adding the reactor prior to the DUNE $\chi^2$ function as,
\begin{equation}
\Delta \chi^2_{\rm Tot}=\Delta \chi^2+\left(\frac{\sin^2\theta_{13}(TEST)-\sin^2\theta_{13}(TRUE)}{\sigma_{13}}\right)^2,
\end{equation}
 where $\sigma_{13}$ is the reactor measured erro on $\theta_{13}$. The result is presented in Fig.~\ref{fig:plot_general}. Where we plotted the exclusion potential regions from 1 to 5 $\sigma$ as a function of the true value of $\theta_{23}$ by assuming $\sin^2\theta_{13}(TRUE)=0.02155$ and $\sigma_{13}=3\%$. The black dashed line is represents the TFSM. We see that for true values of $\theta_{23}$ any such correlation can be excuded around the maximal point $\pi/4$. Nevertheless, too small correlation $f<0.8$ or too strong correlation $f>1.9$ can also excluded at $4\sigma$ for any value of true $\theta_{23}$.  

\begin{figure}[!ht]
\centering
\includegraphics[scale=.4]{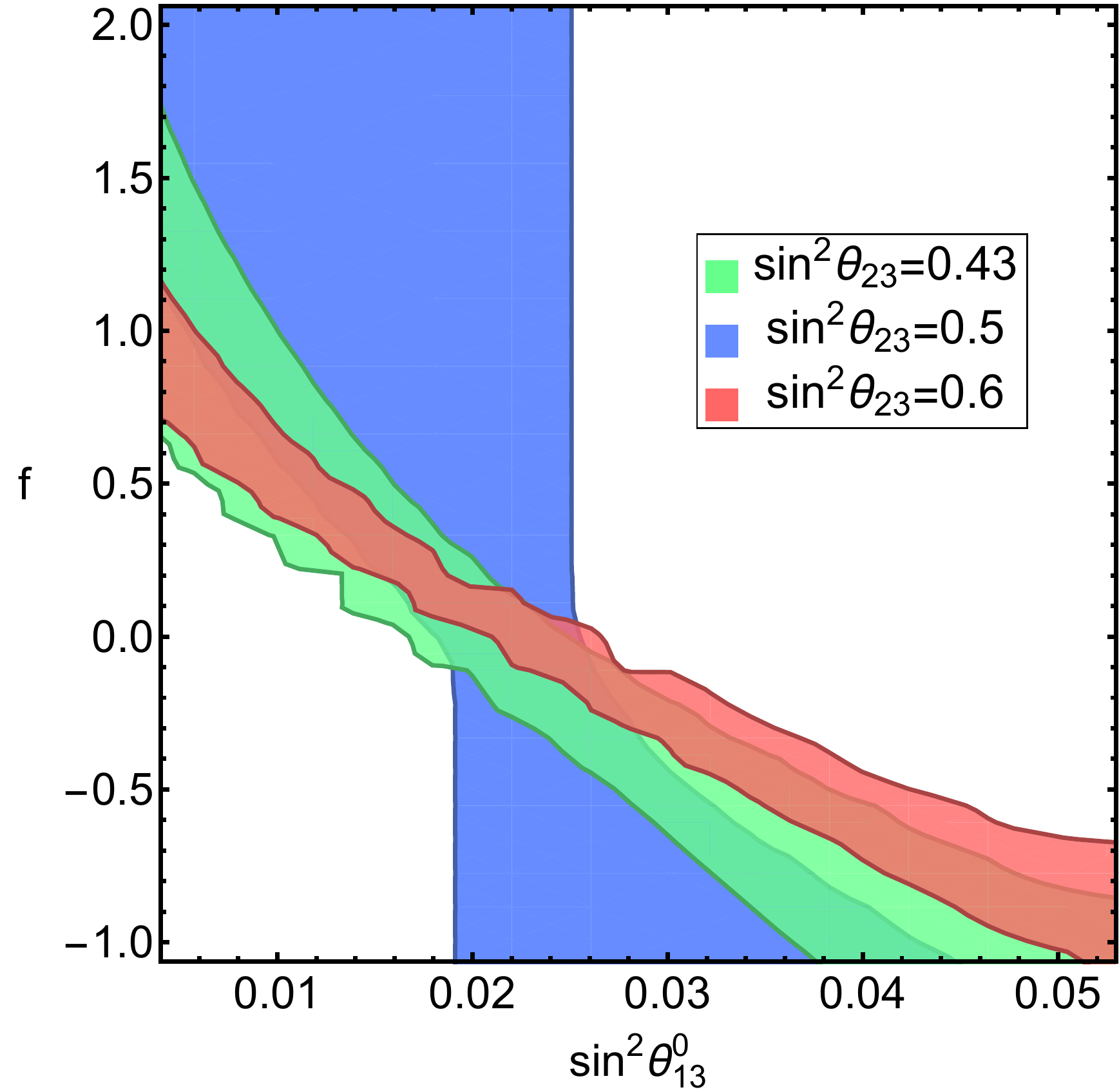}
\caption{\label{fig:plot_general2} $3\sigma$ parameter space $\theta_{13}^0$ and $f$ of relation~\ref{eq:contrain2} that cannot be distinguished from the unconstrained relation hypothesis by DUNE. The true values are: $\sin^2\theta_{13}(TRUE)=0.02155$ and three cases: (1) $\sin^2\theta_{23}(TRUE)=0.43$ (green), (2) $\sin^2\theta_{23}(TRUE)=0.5$ blue and (3) $\sin^2\theta_{23}(TRUE)=0.6$ Red. }
\end{figure}

Finally, in Fig.\ref{fig:plot_general2} we show the $3\sigma$ parameter space $\theta_{13}^0$ and $f$ of relation~\ref{eq:contrain2} that cannot be distinguished from the unconstrained relation hypothesis by DUNE, assuming $\sin^2\theta_{13}(TRUE)=0.02155$ and three cases: (1) $\sin^2\theta_{23}(TRUE)=0.43$ (green), (2) $\sin^2\theta_{23}(TRUE)=0.5$ blue and (3) $\sin^2\theta_{23}(TRUE)=0.6$ Red. Notice that the farther away from the maximality, the smaller the region.

\section{\label{sec:conc} Conclusion}
We showed that it is possible to probe high-energy models that present correlations between the atmospheric and reactor angle by using long-baseline experiments. As an example, we took the Tetrahedral Flavor Symmetry model and simulated its predictions in the DUNE experiment and show that a large portion of the phase space can be excluded. On the other hand, we found that the octant degeneracy creates a $2\sigma$ degeneracy for small $\theta_{13}$ that partially blind the experiment to the model's correlation. Nevertheless, the general correlation $ \theta_{13}=f\left|\frac{\pi}{4}-\theta_{23}\right|$ can be probed if one perform a joint analysis with DUNE and reactors. In special if $f>1.9$ or $f<0.8$ the entire region can be probed if $\sin^2\theta_{13}\sim 0.02155$.
\section*{Acknowledgments}
Pedro Pasquini was supported by FAPESP-CAPES grants 2014/05133-1, 2015/16809-9, 2014/19164-6 and FAEPEX grant N. 2391/17.

 \bibliographystyle{apsrev4-1}
%
\end{document}